\documentclass[fleqn,twoside]{article}
\usepackage{espcrc2}
\usepackage{graphicx}
\usepackage{bm}

\newcommand{\AmS}{{\protect\the\textfont2
  A\kern-.1667em\lower.5ex\hbox{M}\kern-.125emS}}

\title{Group-theoretical construction of extended baryon operators}

\author{LHP Collaboration: 
        S.~Basak\address[UMD]{%
                        Department of Physics, 
                        University of Maryland,
                        College Park, MD 20742, USA},
        R.~Edwards\address[JLAB]{%
                        Thomas Jefferson National Accelerator Facility,
                        Newport News, VA 23606, USA},
        R.~Fiebig\address[FIU]{%
                        Physics Department, 
                        Florida International University, 
                        Miami, FL 33199, USA},
        G.T.~Fleming\address[YALE]{%
                        Sloane Physics Laboratory, 
                        Yale University, 
                        New Haven, CT 06520, USA}%
                        \addressmark[JLAB],
        U.M.~Heller\address{%
                        American Physical Society,
                        Ridge, NY 11961-9000, USA},
        C.~Morningstar\address[CMU]{%
                        Department of Physics, 
                        Carnegie Mellon University, 
                        Pittsburgh, PA 15213, USA}%
                 \thanks{Presented by C.~Morningstar.},
        D.~Richards\addressmark[JLAB],
        I.~Sato\addressmark[UMD],
        S.~Wallace\addressmark[UMD].
        }
       
\begin{document}

\begin{abstract}
The design and implementation of large sets of spatially extended
baryon operators for use in lattice simulations are described.
The operators are constructed to maximize overlaps with the low-lying
states of interest, while minimizing the number of sources needed
in computing the required quark propagators.
%\vspace{-2mm}
\end{abstract}
\maketitle

\section{OVERVIEW}

The Lattice Hadron Physics Collaboration (LHPC) was given a charge
a few years ago by the late Nathan Isgur to extract as much as possible
of the low-lying spectrum of baryon resonances from Monte Carlo simulations
in lattice QCD.  To accomplish this ambitious task, we have been
pursuing two different approaches to constructing baryon interpolating field
operators.  Progress with the first of these approaches is reported
elsewhere in this conference\cite{sato,basak}.  In this talk, developments in the
design and implementation of extended baryon operators using the second
approach are outlined.  Initial efforts using this second method were
described at the Tsukuba symposium last year\cite{lastyear}.  Some details
presented in the earlier reference will not be repeated in this brief report.

The usual operator construction which mimics the approach one would take in
continuous space-time (such as in the QCD sum-rule method) is very cumbersome,
especially when tackling an entire spectrum, as opposed to a single state or
a few states.  Our approach described here is instead to directly combine the
physical characteristics of baryons with the symmetries of the lattice
regularization of QCD used in simulations.  For baryons at rest, our 
operators are formed using group-theoretical projections onto the irreducible
representations (irreps) of the $O_h$ symmetry group of a three-dimensional cubic lattice.
There are four two-dimensional irreps $G_{1g}, G_{1u}, G_{2g}$, $G_{2u}$
and two four-dimensional representations $H_g$ and $H_u$. 
The continuum-limit spins $J$ of our states must be deduced by examining
degeneracy patterns across the different $O_h$ irreps\cite{lastyear}.

Baryons are expected to be rather large objects, and hence, local operators
will not suffice.  Our approach to constructing extended operators is
to use covariant displacements of the quark fields.
Displacements in different directions are used to build up the appropriate
orbital structure, and displacements of different lengths can build up
the needed radial structure.  All of our three-quark baryon operators are
linear superpositions of gauge-invariant terms of the form
\begin{equation}
\Phi^{ABC}_{\alpha i;\beta j;\gamma k}
\!\!=\!    \varepsilon_{abc} (\tilde{D}^{(p)}_i\!\tilde{\psi})^A_{a\alpha} 
   (\tilde{D}^{(p)}_j\!\tilde{\psi})^B_{b\beta}
   (\tilde{D}^{(p)}_k\!\tilde{\psi})^C_{c\gamma},
\end{equation}
where $A,B,C$ indicate quark flavor, $a,b,c$ are color indices,
$\alpha,\beta,\gamma$ are Dirac spin indices,
$\tilde{\psi}$ indicates a smeared quark field, and
$\tilde{D}^{(p)}_j$ denotes the $p$-link covariant displacement
operator in the $j$-th direction. The three-dimensional
gauge-covariant Laplacian is used to smear the quark fields:
\begin{eqnarray}
  \tilde{\psi}(x) &=& (1+\varrho \tilde{\Delta})^{n_\varrho}\ \psi(x),\\
   \tilde{\Delta} \psi(x) &=& \!\!\!\!\!\!\sum_{k=\pm 1, \pm 2,\pm 3} \Bigl(
   \tilde{U}_k(x)\psi(x\!+\!\hat{k})-\psi(x) \Bigr),
\end{eqnarray}
where $\varrho$ and integer $n_\varrho$ are tunable parameters, and
$\tilde{U}_\mu(x)$ denotes a link variable smeared using
the stout link algorithm\cite{stout}.  The $p$-link gauge-covariant
displacement operator is defined by
\begin{equation}
 \tilde{D}_j^{(p)}O(x) = 
  \tilde{U}_j(x)\dots 
   \tilde{U}_j(x\!+\!(p\!-\!1)\hat{j}) O(x\!+\!p\hat{j}).
\end{equation}
In the definition of $\Phi^{ABC}_{\alpha i;\beta j;\gamma k}$ above,
the displacement directions $i,j,k$ can take values $0,\pm 1,\pm 2,\pm 3$,
where the value $0$ indicates that no displacement occurs.  To simplify
matters, the displacement length $p$ is taken to be the same for all
displaced quarks in a given term.  Ultimately, there are six different
spatial orientations that we use, shown in Table~\ref{tab:opforms}.  The
singly-displaced operators are meant to mock up a diquark-quark coupling,
and the doubly-displaced and triply-displaced operators are chosen since
they favor the $\Delta$-flux and $Y$-flux configurations, respectively.

\begin{table}[th]
\caption[captab]{Six types of three-quark 
$\Phi^{ABC}_{\alpha i;\beta j;\gamma k}$ operators used.
Smeared quarks fields are
shown by solid circles, line segments indicate
covariant displacements, and each hollow circle indicates the location
of a color $\varepsilon_{abc}$ coupling.  For simplicity, all displacements
have the same length in an operator. 
\label{tab:opforms}}
\begin{center}
\begin{tabular}{c@{\hspace{3mm}}l}\hline
 Operator type &  Displacement indices\\ \hline
\raisebox{0mm}{\setlength{\unitlength}{1mm}
\thicklines
\begin{picture}(16,10)
\put(8,6.5){\circle{6}}
\put(7,6){\circle*{2}}
\put(9,6){\circle*{2}}
\put(8,8){\circle*{2}}
\put(0,0){single-site}
\end{picture}}  & \raisebox{3mm}{$i=j=k=0$ }\\ 
\raisebox{0mm}{\setlength{\unitlength}{1mm}
\thicklines
\begin{picture}(23,10)
\put(7,6.2){\circle{5}}
\put(7,5){\circle*{2}}
\put(7,7.3){\circle*{2}}
\put(14,6){\circle*{2}}
\put(9.5,6){\line(1,0){4}}
\put(0,0){singly-displaced}
\end{picture}}  & \raisebox{3mm}{$i=j=0,\ k\neq 0$} \\ 
\raisebox{0mm}{\setlength{\unitlength}{1mm}
\thicklines
\begin{picture}(26,8)
\put(12,5){\circle{3}}
\put(12,5){\circle*{2}}
\put(6,5){\circle*{2}}
\put(18,5){\circle*{2}}
\put(6,5){\line(1,0){4.2}}
\put(18,5){\line(-1,0){4.2}}
\put(-1,0){doubly-displaced-I}
\end{picture}}  & \raisebox{2mm}{$i=0,\ j=-k,\ k\neq 0$} \\ 
\raisebox{0mm}{\setlength{\unitlength}{1mm}
\thicklines
\begin{picture}(20,13)
\put(8,5){\circle{3}}
\put(8,5){\circle*{2}}
\put(8,11){\circle*{2}}
\put(14,5){\circle*{2}}
\put(14,5){\line(-1,0){4.2}}
\put(8,11){\line(0,-1){4.2}}
\put(-5,0){doubly-displaced-L}
\end{picture}}   & \raisebox{4mm}{$i=0,\ \vert j\vert\neq \vert k\vert,
  \ jk\neq 0$}\\ 
\raisebox{0mm}{\setlength{\unitlength}{1mm}
\thicklines
\begin{picture}(20,12)
\put(10,10){\circle{2}}
\put(4,10){\circle*{2}}
\put(16,10){\circle*{2}}
\put(10,4){\circle*{2}}
\put(4,10){\line(1,0){5}}
\put(16,10){\line(-1,0){5}}
\put(10,4){\line(0,1){5}}
\put(-5,0){triply-displaced-T}
\end{picture}}   & \raisebox{4mm}{$i=-j,\ \vert j\vert \neq\vert k\vert,
 \ jk\neq 0$} \\ 
\raisebox{0mm}{\setlength{\unitlength}{1mm}
\thicklines
\begin{picture}(20,12)
\put(10,10){\circle{2}}
\put(6,6){\circle*{2}}
\put(16,10){\circle*{2}}
\put(10,4){\circle*{2}}
\put(6,6){\line(1,1){3.6}}
\put(16,10){\line(-1,0){5}}
\put(10,4){\line(0,1){5}}
\put(-5,0){triply-displaced-O}
\end{picture}}   & \raisebox{4mm}{$\vert i\vert \neq \vert j\vert \neq
  \vert k\vert,\ ijk\neq 0$} \\ \hline
\end{tabular}
\end{center}
\end{table}

Next, we combine the $\Phi^{ABC}_{\alpha i;\beta j;\gamma k}$ into 
{\em elemental} operators $B^{F}_i(t,\bm{x})$ having the appropriate flavor
structure characterized by isospin, strangeness, {\it etc.}   We work in 
the $m_u=m_d$ (equal $u$ and $d$ quark masses) approximation, and thus,
require that the elemental operators have definite isospin, that is, they 
satisfy appropriate commutation relations with the isospin operators
$\tau_3,\tau_+,\tau_-$.  Since we plan to compute full correlation matrices,
we need not be concerned with forming operators according to an $SU(3)$ 
flavor symmetry.  For zero momentum states, we impose
translation invariance:  $B_i^F(t)=\sum_{\bm{x}} B_i^F(t,\bm{x})$.
Maple code which manipulates Grassmann fields was used to identify
maximal sets of linearly independent elemental operators.
The numbers of such operators are listed in Table~\ref{tab:elemental}.

\begin{table}
\caption{Numbers of linearly independent elemental operators.
$\Delta^{++}$ operators have $uuu$ flavor structure,
 ($\Omega^-$ have $sss$),
$N^+$ operators are $uud-duu$,
$\Sigma^+$ are $uus$ and $suu$
 ($\Xi^0$ are $ssu$ and $uss$), and 
$\Lambda^0$ operators have $uds-sud$ flavor structure.
\label{tab:elemental}}
\begin{center}
\begin{tabular}{crrrr}\hline
Operator Type & $\Delta,\Omega$ & $N$ & $\Sigma,\Xi$ 
   & $\Lambda$\\ \hline
 single-site        &  20  & 20  &  40   & 24  \\
 singly-displaced   &  240 & 384 &  624  & 528 \\
\hspace{-3mm} doubly-displaced-I &  192 & 384 &  576  & 576 \\
\hspace{-3mm} doubly-displaced-L &  768 & 1536&  2304 & 2304\\
 triply-displaced-T &  768 & 1536&  2304 & 2304\\
 triply-displaced-O &  512 & 1024&  1536 & 1536\\ \hline
\end{tabular}
\end{center}
\end{table}

The final step in our operator construction is to
apply group-theoretical projections to
obtain operators which transform irreducibly under 
all lattice rotation and reflection symmetries:
\begin{equation}
  B_i^{\Lambda\lambda F}\!(t)\!
 = \!\frac{d_\Lambda}{g_{O_h}}\!\!\!\textstyle\sum_{R\in O_h}
  \!\!D^{(\Lambda)}_{\lambda\lambda}(R)^\ast
   U_R B^F_i\!(t) U_R^\dagger,
\label{eq:project}\end{equation}
where $\Lambda$ refers to an $O_h$ irrep, $\lambda$ is the irrep row,
$g_{O_h}$ is the number of elements in $O_h$,
$d_\Lambda$ is the dimension of the $\Lambda$ irrep,
$D^{(\Lambda)}_{mn}(R)$ is a $\Lambda$
representation matrix corresponding to group element $R$,
and $U_R$ is the quantum operator which implements the symmetry
operations. 

The projections in Eq.~(\ref{eq:project}) are carried out
using Maple in the following sequence of steps.
(a) A set of $M_B$ linearly-independent elemental operators
$B_j^F(t)$ that transform among one another under $O_h$ is identified.
(b) We obtain the $M_B\times M_B$ representation matrices 
$D_{ij}(R)$ which satisfy
\begin{equation}
  U_R\ B_i^F(t)\ U_R^\dagger = \textstyle\sum_{j=1}^{M_B} B_j^F(t)\ D_{ji}(R).
\end{equation}
(c) Since the resulting representations may not be unitary,
we need to compute the Hermitian metric matrix $M$
\begin{equation}
 M_{ij} = \frac{1}{g_{O_h^D}}\textstyle\sum_{R\in O_h^D} \sum_{k=1}^{M_B}
   D_{ki}(R)^\ast D_{kj}(R).
\end{equation}
(d) For each irrep $\Lambda$, we compute the $M_B\!\times\! M_B$
projection matrix for row $\lambda=1$:
\begin{equation}
  P^{\Lambda\lambda F}_{ij}\! =\!
 \frac{d_\Lambda}{g_{O_h^D}}\textstyle\sum_{R\in O_h^D} 
  \left[D^{(\Lambda)}_{\lambda\lambda}(R)^\ast D_{ji}(R)
 \right]_{\lambda\!=\!1}.
\end{equation}
(e) From the rows of this projection matrix, 
$r$ linearly-independent operators are obtained: 
\begin{eqnarray}
 &B^{\Lambda\lambda F}_i(t) = \textstyle\sum_{j=1}^{M_B}\ c^{\Lambda\lambda F}_{ij}
 \ B_j^F(t), \quad  (\lambda=1)&\\
 &\textstyle\sum_{k,l=1}^{M_B} c^{\Lambda\lambda F\ast}_{ik}
 \ M_{kl}\ c^{\Lambda\lambda F}_{jl} = \delta_{ij},
 \ (i=1\!\dots \!r).&
\end{eqnarray}
(f) For each of the $r$ operators $B^{\Lambda\lambda F}_i(t)$ in
the first row  $\lambda=1$, we
obtain partner operators in all other rows $\mu>1$ using
\begin{equation}
 c^{\Lambda\mu F}_{ik}
  = \sum_{j=1}^{M_B} c^{\Lambda\lambda F}_{ij}
  \frac{d_\Lambda}{g_{O_h^D}}\sum_{R\in O_h^D} 
  D^{(\Lambda)}_{\mu\lambda}(R)^\ast
  \ D_{kj}(R).
\end{equation}
We average over rows for increased statistics.  The total numbers of
operators one obtains in each row in the different isospin channels are
given in Table~\ref{tab:totalnumbers}.

\begin{table}
\caption{Total numbers of baryon operators for
two different displacement lengths.\label{tab:totalnumbers}}
\begin{center}
\begin{tabular}{crrrr}\hline
  Irrep & $\Delta,\Omega$ & $N$ & $\Sigma,\Xi$ 
   & $\Lambda$\\ \hline
 $G_{1g}$  &  221 & 443&  664  & 656 \\
 $G_{1u}$  &  221 & 443&  664  & 656 \\
 $G_{2g}$  &  188 & 376&  564  & 556 \\
 $G_{2u}$  &  188 & 376&  564  & 556 \\
 $H_g$     &  418 & 809&  1227 & 1209\\
 $H_u$     &  418 & 809&  1227 & 1209\\ \hline
\end{tabular}
\end{center}
\end{table}

With such large numbers of operators, minimizing the number of
quark propagator sources is crucial.  We have recently finished
writing software which, for each element of the baryon correlation
matrices, performs all of the needed Wick contractions and
simplifies the results by rotating individual terms to
minimize the number of displacement orientations of the baryon
sources.  We hope to report on initial quenched simulations
in the near future.  Due to the large sizes of the baryons
and their large masses, the use of improved actions on anisotropic
lattices is important.  Ultimately, when quark loops are included at
realistically light quark masses, multi-hadron (baryon-meson) 
operators must be included in our correlation matrices, and finite-volume
techniques will need to be employed to ferret out the baryon resonances
from uninteresting scattering states.  We are currently exploring
different ways of building such operators. 

Although we focused on tri-quark baryon operators, our approach
to constructing hadronic operators is applicable to mesons, tetra-quark,
and penta-quark operators.  By including $\bm{E}$ and $\bm{B}$ fields
into our operators, hadrons bound by an excited gluon field
can also be studied.
This work was supported by the U.S.~National Science Foundation 
through grants PHY-0099450 and PHY-0300065, and by 
the U.S.~Department of Energy under
contracts DE-AC05-84ER40150 and DE-FG02-93ER-40762.

\end{document}